\documentstyle[12pt]{article}
\input amssym.def
\input amssym.tex

\newcommand{\g}{{\rm\bf g}}
\newcommand{\uu}{{\rm\bf u}}
\newcommand{\vv}{{\rm\bf v}}

\newcommand{\A}{{\rm\bf A}}
\newcommand{\B}{{\rm\bf B}}
\newcommand{\C}{{\rm\bf C}}
\newcommand{\D}{{\rm\bf D}}

\newcommand{\qed}{\quad\rule{3mm}{3mm}}

\newtheorem{proposition}{Proposition}
\newtheorem{theorem}{Theorem}

\begin{document}

\begin{titlepage}{\LARGE
\begin{center} Nonlocal quadratic Poisson algebras, \\
 monodromy map,\\ and Bogoyavlensky lattices
 \end{center}} \vspace{1.5cm}
\begin{flushleft}{\large Yuri B. SURIS}\end{flushleft} \vspace{1.0cm}
Centre for Complex Systems and Visualization, University of Bremen,\\
Universit\"atsallee 29, 28359 Bremen, Germany\\
e-mail: suris @ cevis.uni-bremen.de

\vspace{1.5cm}  
{\small {\bf Abstract.} A new Lax representation for the Bogoyavlensky
lattice is found, its $r$--matrix interpretation is elaborated. The
$r$--matrix structure turns out to be related to a highly nonlocal
quadratic Poisson structure on a direct sum of associative algebras.
The theory of such nonlocal structures is developed, the Poisson
property of the monodromy map is worked out in the most general 
situation. Some problems concerning the duality of Lax representations
are raised. }
 
\end{titlepage}

\setcounter{equation}{0}
\section{Introduction}
In this work we find some new results on the well known integrable 
system, the Bogoyavlensky lattice. However, these results, dealing with one
particular system, allow to touch two general problems of the Hamiltonian
theory of integrable systems.

The first of these problems is the nature of {\it the most general quadratic
Poisson brackets} on associative algebras (or, in a different setting, on Lie 
groups). These brackets were invented and used for various purposes
in \cite{FM}, \cite{P}, \cite{S}, \cite{STS3}, \cite{STS4}, \cite{STSS},
\cite{O}. They serve as a wide generalization of the so called Sklyanin
bracket \cite{STS1}, \cite{RSTS}. The characteristic property  of the Sklyanin
bracket in the Lie groups setting is that it defines a Lie--Poisson structure,
i.e. the group multiplication is a Poisson map with respect to this bracket.
In the framework of associative algebras the corresponding property
may be formulated in the following manner. Let the algebra $g$ carry the 
Sklyanin bracket, consider the direct sum $\g=g^{(1)}\oplus\ldots\oplus g^{(n)}$ 
($g^{(k)}$ are $n$ copies of $g$), and supply $\g$ with
the direct sum of the Sklyanin brackets. Then the monodromy map $\g\mapsto g$,
defined as $(u_1,\ldots,u_n)\mapsto u_n\cdot\ldots\cdot u_1$, has the Poisson
property. Although several generalizations of this statement are available in 
the literature, the most general version still has not appeared, presumably 
because of the lack of interesting examples.

The most examples where the monodomy map is of interest, are connected with
the lattice models, where each $g^{(k)}$ in the direct sum $\g$ is attached 
to the $k$th lattice site and plays the role of the phase space of the $k$th
particle. In all known examples the Poisson bracket on $\g$ is
either ultralocal (the functions depending only on the $g^{(k)}$ are in
involution for different $k$), or non--ultralocal, but still with some 
locality properties (usually the functions on $g^{(k)}$ are in 
involution with the functions on $g^{(j)}$, unless $|j-k|\le 1$). 

In this paper we establish the Poisson property of the monodromy map in 
the most general situation (see Proposition 5 and Theorem 1). Namely, we 
allow an arbitrary nonlocality, when the functions on $g^{(k)}$ do not commute 
with the functions on $g^{(j)}$, irrespectively of the value of $|j-k|$.
We give also a rather spectacular example of the situation where our 
construction works (and is natural and necessary) -- the well known 
Bogoyavlensky lattice. It is interesting to note that in this example
different $g^{(k)}$'s are attached not to single lattice sites, but to sets of
equidistant sites (or, in other terminology, to different sorts of particles).

Not very surprisingly, to obtain these new results for an old system,
we have to deal with a novel Lax representation for it. Here we touch the
second general problem: the {\it duality}. It is well known that often
one and the same integrable system admits two (or more) Lax representations,
living in different Lie algebras. In particular, the Lax matrices of
these representations have different dimensions. 

If one knows a Lax representation for a system at hand, then it is often 
possible to find an alternative Lax representation by the  following trick.
Take a characteristic polynomial of the known Lax matrix. It serves as 
a generating function for the integrals of motion. Using some determinantal
identities, one can often represent this function as a characteristic
polynomial of some other matrix. Then there is a good chance that this matrix
is indeed a (new) Lax matrix, i.e. that it takes part in a
Lax representation  for our system. 

One of the determinantal identities often used by such transformations, is
the so called Weinstein--Aronszajn formula \cite{M}, \cite{AHH}. A novel
feature of the present work is that instead of the latter formula we use the 
Laplace expansion in order to find new Lax matrices.

What is however very important and not provided by determinantal identities,
is establishing a correspondence (if possible, possessing certain Poisson 
properties) between different Lax matrices, rather then between their 
characteristic polynomials. To my knowledge, it was done only in the framework 
of linear Poisson brackets \cite{AHH}. It would be highly desirable to work out 
such correspondences in various different contexts, e.g., if one of the Lax 
matrices belongs to a linear and another to a quadratic Poisson algebra (the 
most prominent example being the usual Toda lattice, cf. \cite{RSTS}, 
\cite{FT}), or if the both matrices belong to quadratic Poisson algebras (e.g., 
the relativistic Toda lattice, cf. \cite{S1},\cite{S2},\cite{S}). 

The example elaborated in the present paper belongs to the latter class (both 
algebras quadratic), and has two additional interesting features. First, the 
Poisson structure of the first (old) Lax formulation includes a Dirac reduction, 
and, second, the dimensions of the both Lax matrices (old and new) grow with 
the number of particles.

\setcounter{equation}{0}
\section{Quadratic brackets on associative algebras}

Let $g$ be an assosiative algebra equipped with a nondegenerate scalar product
$\langle\cdot,\cdot\rangle$ which is invariant in the following sense:
\begin{equation}
\langle u\cdot v,w\rangle=\langle u,v\cdot w\rangle.
\end{equation}
The gradient $\nabla\varphi\in g$ of a smooth function $\varphi$ on $g$ is  
defined by the following relation:
\begin{equation}
\langle\nabla\varphi(u),X\rangle=\left.\frac{d}{d\varepsilon}
\varphi(u+\varepsilon X)\right|_{\varepsilon=0}\qquad\forall X\in g.
\end{equation}
Denote also
\begin{equation}
d\varphi(u)=u\cdot\nabla\varphi(u),\qquad d\,'\varphi(u)=\nabla\varphi(u)\cdot u.
\end{equation}

The following ''hierarchy'' of quadratic brackets on $g$ is known.

1) The Sklyanin bracket \cite{STS1}, \cite{STS2}:
\[
\{\varphi,\psi\}(u)=\langle R(d\,'\varphi),d\,'\psi\rangle -
\langle R(d\varphi),d\psi\rangle.
\]
The linear operator $R$ on $g$ has to be skew--symmetric:
\[
R^*=-R
\]
in order to assure the skew--symmetry of the bracket; here $ ^*$ denotes the 
adjoint operator with respect to the scalar product
$\langle\cdot,\cdot\rangle$. A sufficient condition
for this bracket to satisfy the Jacobi identity is the so called
modified Yang--Baxter equation for the operator $R$:
\begin{equation}\label{mYB}
[R(X),R(Y)]=R\Big([R(X),Y]+[X,R(Y)]\Big)-\alpha[X,Y]\qquad \forall X,Y\in g
\end{equation}
with some constant $\alpha$. We shall denote this condition by mYB($R;\alpha$).

2) The Li--Parmentier--Oevel--Ragnisco bracket \cite{LP1}, \cite{OR}:
\[
\{\varphi,\psi\}(u)  = \langle A(d\,'\varphi),\,d\,'\psi\rangle-
\langle A(d\varphi),\,d\psi\rangle
+\langle S(d\varphi),\,d\,'\psi\rangle-\langle S(d\,'\varphi),\,d\psi\rangle.
\]
For the skew--symmetry of this bracket one needs
\[
A^*=-A,\qquad S^*=S.
\]
A sufficient condition for the Jacobi identity is given by mYB($A;\alpha$) and
an additional property of the linear operators $A$, $S$:
\begin{equation}\label{Hom}
[S(X),S(Y)]=S\Big([A(X),Y]+[X,A(Y)]\Big)\qquad \forall X,Y\in g.
\end{equation}
We shall denote this condition by Hom($S,A$). It turns out that the set of 
two conditions mYB($A;\alpha$), Hom($S,A$) is equivalent also to the set of
two conditions mYB($A;\alpha$), mYB($A+S;\alpha$).

3) The most general quadratic bracket on $g$ \cite{FM}, \cite{P}, \cite{S}:
\begin{equation}\label{q br}
\{\varphi,\psi\}(u)  = \langle A(d\,'\varphi),\,d\,'\psi\rangle-
\langle D(d\varphi),\,d\psi\rangle
+\langle B(d\varphi),\,d\,'\psi\rangle-\langle C(d\,'\varphi),\,d\psi\rangle.
\end{equation}
We shall denote this expression by PB($A, B, C, D$). 
For the skew--symmetry one needs the following conditions on the linear 
operators $A, B, C, D$:
\begin{equation}\label{skew}
A^*=-A,\qquad D^*=-D,\qquad B^*=C.
\end{equation}

\begin{proposition} 
Sufficient conditions for {\rm (\ref{q br})} to be
a Poisson bracket are given by the conditions 
\begin{equation}\label{cond1}
{\rm mYB(}A;\alpha), \quad {\rm mYB(}D;\alpha), \quad
{\rm Hom(}B,D),\quad {\rm Hom(}C,A). 
\end{equation}
If
\begin{equation}\label{sum}
A+B=C+D=R,
\end{equation}
then the above sufficient conditions are equivalent to 
\begin{equation}\label{cond2}
{\rm mYB(}A;\alpha),\quad {\rm mYB(}D;\alpha),\quad {\rm mYB(}R;\alpha).
\end{equation}
\end{proposition}

Under conditions given in the Proposition 1, following Hamiltonian system
on $g$ is defined for every smooth function $\varphi$:
\begin{eqnarray}
\dot{u}=\{\varphi,u\} & = &
u\cdot A(d\,'\varphi(u))-D(d\varphi(u))\cdot u \nonumber\\
                        &   &
+u\cdot B(d\varphi(u))-C(d\,'\varphi(u))\cdot u . 
\label{eq of motion}
\end{eqnarray}
If (\ref{sum}) is fulfilled, then this equation takes the Lax form for 
$Ad$--invariant Hamiltonian functions $\varphi$: for such functions
\begin{equation}
d\varphi(u)=d\,'\varphi(u)
\end{equation}
and we get the equation in the Lax form :
\begin{equation}\label{Lax q}
\dot{u}=\{\varphi,u\}=[u,\,R(d\varphi(u))].
\end{equation}

\begin{proposition} 
Under the condition {\rm(\ref{sum})} $Ad$--invariant 
functions are in involution with respect to the bracket $\{\cdot,\cdot\}$, 
hence each $Ad$--invariant function is an integral of motion for 
{\rm (\ref{Lax q})}.
\end{proposition}

\setcounter{equation}{0}
\section{Poisson brackets on direct sums}
Consider a ''big'' algebra $\g=\bigoplus_{k=1}^n g$, a direct sum of $n$
copies of the algebra $g$. So, the multiplication in $\g$ is componentwise:
if $\uu=(u_1,\ldots,u_n)\in\g$, $\vv=(v_1,\ldots,v_n)\in\g$, then
\[
\uu\cdot\vv=(u_1\cdot v_1,\ldots,u_n\cdot v_n).
\]
Define the (nondegenerate, invariant) scalar product 
$\langle\langle\cdot,\cdot\rangle\rangle$ on $\g$ as
\[
\langle\langle\uu,\vv\rangle\rangle=\sum_{k=1}^n\langle u_k,v_k\rangle.
\]

Now let $\A, \B, \C, \D$ be linear operators on $\g$ satisfying conditions
analogous to (\ref{skew}) and to (\ref{cond1}) (or (\ref{cond2})). Then
one can define the bracket PB($\A, \B, \C, \D$) on $\g$:
\begin{eqnarray}
\{\Phi,\Psi\}(\uu) & = & \langle\langle \A(d\,'\Phi),\,d\,'\Psi\rangle\rangle-
\langle\langle \D(d\Phi),\,d\Psi\rangle\rangle  \nonumber\\
 & + & \langle\langle \B(d\Phi),\,d\,'\Psi\rangle\rangle
-\langle\langle \C(d\,'\Phi),\,d\Psi\rangle\rangle.
\label{big br}
\end{eqnarray}
To be able to deal
with these objects, let us introduce following (natural) notations. Let
$\A_i(\uu)$ be the $i$th component of $\A(\uu)$; then we set
\begin{equation}\label{A i}
\A_i(\uu)=\sum_{j=1}^n A_{ij}(u_j).
\end{equation}
Let also  $\nabla_j\Phi$, $d_j\Phi$, $d\,'_{\!j}\Phi$ denote the $j$th 
components of the corresponding objects for a smooth function $\Phi(\uu)$
on $\g$.

Then the conditions (\ref{skew}) for the operators $\A, \B, \C, \D$ read:
\[
A_{ij}^*=-A_{ji},\qquad D_{ij}^*=-D_{ji},\qquad B_{ij}^*=C_{ji}.
\]
The bracket (\ref{big br}) takes the form
\begin{eqnarray}\label{big PB}
\{\Phi,\Psi\}(\uu) & = &
\sum_{i,j=1}^n\langle A_{ij}(d\,'_{\!j}\Phi),d\,'_{\!i}\Psi\rangle-
\sum_{i,j=1}^n\langle D_{ij}(d_j\Phi),d_i\Psi\rangle\nonumber\\
 & & +\sum_{i,j=1}^n\langle B_{ij}(d_j\Phi),d\,'_{\!i}\Psi\rangle
-\sum_{i,j=1}^n\langle C_{ij}(d\,'_{\!j}\Phi),d_i\Psi\rangle
\end{eqnarray}

It is interesting to reformulate the conditions of the Proposition 1 for
the operators $\A, \B, \C, \D$ acting on $\g$ in terms of the operators
$A_{ij}$, $B_{ij}$, $C_{ij}$, $D_{ij}$ acting on $g$.
\begin{proposition} 
The condition {\rm mYB($\A;\alpha$)} for a skew symmetric operator $\A$
is equivalent to the following set of conditions:

1) $n$ equations {\rm mYB($A_{jj};\alpha$)};

2) $n(n-1)$ equations {\rm Hom($A_{ij}, A_{jj}$)} (for all $i\neq j$);

3) $n(n-1)(n-2)/6$ conditions (for all $i<j<k$):
\[
[A_{ij}(X),A_{ik}(Y)]=A_{ik}\Big([A_{kj}(X),Y]\Big)
+A_{ij}\Big([X,A_{jk}(Y)]\Big) \qquad \forall X,Y\in g.
\]
\end{proposition}

We shall denote the last condition by Aux($A_{ij}, A_{ik}; A_{kj}, A_{jk}$).

\begin{proposition}
The condition {\rm Hom($\C, \A$)} is equivalent to the set of 

1) $n^2$ conditions {\rm Hom($C_{ij}, A_{jj}$)} (for all $1\leq i,j\leq n$);

2) $n^2(n-1)/2$ conditions {\rm Aux($C_{ij}, C_{ik}; A_{kj}, A_{jk}$)} 
(for all $1\le i,j,k\le n$, $j<k$).
\end{proposition}

{\bf Remark 1.} 
Let us comment, why in the Proposition 3 it is enough to require 
Aux($A_{ij}, A_{ik}; A_{kj}, A_{jk}$) only for $i<j<k$. First, this condition 
is obviously (skew)symmetric with respect to the interchange 
$j\leftrightarrow k$.
Further, it turns out to be symmetric also with respect to cyclic shifts
of the triple $(i,j,k)$ of distinct indices. To demonstrate this, notice that 
due to nondegeneracy of the scalar product the above condition may be expressed 
as
\[
\Big\langle [A_{ij}(X),A_{ik}(Y)],Z\Big\rangle=
\Big\langle A_{ik}\Big([A_{kj}(X),Y]\Big),Z\Big\rangle
+\Big\langle A_{ij}\Big([X,A_{jk}(Y)]\Big),Z\Big\rangle
\]
for arbitrary $X,Y,Z\in g$. Using the skew--symmetry of the operator $\A$ 
($A^*_{ij}=-A_{ji}$) and invariance of the scalar
product, we can transform this into
\[
-\Big\langle X,A_{ji}\Big([A_{ik}(Y),Z]\Big)\Big\rangle=
\Big\langle X,A_{jk}\Big([Y,A_{ki}(Z)]\Big)\Big\rangle
-\Big\langle X,[A_{jk}(Y),A_{ji}(Z)]\Big\rangle
\]
for arbitrary $X,Y,Z\in g$. Again, due to nondegeneracy of the scalar product 
this is equivalent to
\[
-A_{ji}\Big([A_{ik}(Y),Z]\Big)=
A_{jk}\Big([Y,A_{ki}(Z)]\Big)
-[A_{jk}(Y),A_{ji}(Z)]
\]
for arbitrary $Y,Z\in g$. This is exactly Aux($A_{jk}, A_{ji}; A_{ik}, A_{ki}$),
i.e. the above condition for the triple $(j,k,i)$.

{\bf Remark 2.} 
For $n=2$ we see that the mYB($\A;\alpha$) for the skew symmetric operator
\[
\A=\left(\begin{array}{cc} A_{11} & A_{12}\\ A_{21} & A_{22}\end{array}\right)
\]
on $g\bigoplus g$ is equivalent to the set of four conditions
\[
{\rm mYB(}A_{11};\alpha),\quad {\rm mYB(}A_{22};\alpha), \quad
{\rm Hom(}A_{12}, A_{22}),\quad  {\rm Hom(}A_{21}, A_{11}) 
\]
for the operators on $g$.
Comparing with the Proposition 1, we see that it is exactly the sufficient
conditions for PB($A_{11}, A_{12}, A_{21}, A_{22}$) to be a Poisson bracket. 
This coincidence was first pointed out in \cite{STS3}.

\setcounter{equation}{0}
\section{Poisson properties of the monodromy map}

An interesting and important question arising in connection with the Poisson
brackets on the direct sums is the question on the Poisson properties of the
monodromy maps.

\begin{proposition}
Let $\g$ be equipped with the Poisson bracket {\rm PB($\A, \B, \C, \D$)}.
Suppose that the following relations hold:
\[
A_{i+1,j+1}-D_{i,j}+B_{i+1,j}-C_{i,j+1}=0\quad {\rm for}\quad 1\leq i,j\leq n-1;
\]
\[
A_{1,j+1}+B_{1,j}=0\Longleftrightarrow A_{j+1,1}-C_{j,1}=0\quad {\rm for}\;\;
1\leq j\leq n-1;
\]
\[
D_{n,j}+C_{n,j+1}=0\Longleftrightarrow D_{j,n}-B_{j+1,n}=0\quad {\rm for}\;\;
1\leq j\leq n-1.
\]
Then the map 
\begin{equation}\label{monodromy}
{\cal M}: \g\mapsto g,\qquad {\cal M}(\uu)={\cal M}(u_1,\ldots,u_n)=
u_n\cdot\ldots\cdot u_1
\end{equation}
is Poisson, if $g$ is equipped with 
{\rm PB($A_{11}, B_{1n}, C_{n1}, D_{nn}$)}.

\end{proposition}
{\bf Proof.} Take two smooth functions $\varphi$, $\psi$ on $g$, and form two
functions $\Phi$, $\Psi$ on $\g$ according to
\[
\Phi(\uu)=\varphi(T),\quad \Psi(\uu)=\psi(T),
\]
where
\[
T={\cal M}(\uu)=u_n\cdot\ldots\cdot u_1.
\]
It is easy to see that
\[
\nabla_j\Phi(\uu)=u_{j-1}\cdot\ldots\cdot u_1\cdot\nabla\varphi(T)\cdot
u_n\cdot\ldots\cdot u_{j+1}.
\] 
Consequently,
\begin{equation}\label{dPhi}
d_j\Phi(\uu)=u_j\cdot\nabla_j\Phi(\uu)=
u_j\cdot\ldots\cdot u_1\cdot\nabla\varphi(T)\cdot u_n\cdot\ldots\cdot u_{j+1},
\end{equation}
\begin{equation}\label{d'Phi}
d\,'_{\!j}\Phi(\uu)=\nabla_j\Phi(\uu)\cdot u_j=
u_{j-1}\cdot\ldots\cdot u_1\cdot\nabla\varphi(T)\cdot u_n\cdot\ldots\cdot u_j.
\end{equation}
In particular,
\[
d_n\Phi(\uu)=d\varphi(T),\qquad d\,'_{\!1}\Phi(\uu)=d\,'\varphi(T),
\]
and for $1\leq j\leq n-1$ we have
\[
d_j\Phi(\uu)=d\,'_{\!j+1}\Phi(\uu).
\]
Substituting the last two formulas into (\ref{big PB}), we see that under
the conditions of the Proposition, almost all the terms cancel, leaving us with
\begin{eqnarray*}
\{\Phi,\Psi\}(\uu) & = & \langle A_{11}(d\,'\varphi(T)),d\,'\psi(T)\rangle
-\langle D_{nn}(d\varphi(T)),d\psi(T)\rangle\\
 & & +\langle B_{1n}(d\varphi(T)),d\,'\psi(T)\rangle
-\langle C_{n1}(d\,'\varphi(T)),d\psi(T)\rangle,
\end{eqnarray*}
which proves the Proposition.\qed

A further important observation is related to the form of the Lax equations
on $\g$ in the case when the Hamiltonian function has the form
\begin{equation}\label{inv ham}
\Phi(\uu)=\varphi({\cal M}(\uu))=\varphi(u_n\cdot\ldots\cdot u_1), 
\end{equation}
and $\varphi$ is an $Ad$--invariant function on $g$. Under this condition we
can represent the formulas (\ref{dPhi}), (\ref{d'Phi}) as
\begin{equation}\label{dd'Phi}
d_j\Phi(\uu)=d\varphi(T_j),\qquad d\,'_{\!j}\Phi(\uu)=d\varphi(T_{j-1}),
\end{equation}
where
\[
T_j=u_j\cdot\ldots\cdot u_1\cdot u_n\cdot\ldots\cdot u_{j+1}
\]
(so that, in particular, $T_0=T_n=T={\cal M}(\uu)$).
\begin{proposition}
Let $\g$ be equipped with the Poisson bracket {\rm PB($\A, \B, \C, \D$)}.
Then the Hamiltonian equations of motion generated by the Hamiltonian
function {\rm(\ref{inv ham})} with an $Ad$--invariant function $\varphi$, may be
presented in the form
\begin{equation}
\dot{u}_i=u_i{\cal R}_i-{\cal L}_iu_i,
\end{equation}
where
\[
{\cal R}_i=\sum_{j=1}^n (A_{i,j+1}+B_{i,j})(d\varphi(T_j)),
\]
\[
{\cal L}_i=\sum_{j=1}^n (D_{i,j}+C_{i,j+1})(d\varphi(T_j)),
\]
(in these formulas the subscripts should be taken {\rm (mod $n$)}, so that
$A_{i,n+1}=A_{i,1}$ and $C_{i,n+1}=C_{i,1}$).
\end{proposition}
{\bf Proof.} This follows immediately from the equations of motion on $\g$
analogous to (\ref{eq of motion}), the notation (\ref{A i}), and the formulas
(\ref{dd'Phi}).\qed

\vspace{5mm}

The most important and interesting for applications is the situation, when not
only the monodromy map $\cal M$ is Poisson, but also its compositions with
the different powers of the shift
\begin{equation}\label{shift}
\sigma:\g\mapsto\g,\qquad 
\sigma(u_1,\ldots,u_{n-1},u_n)=(u_2,\ldots,u_{n},u_1).
\end{equation}
Shifting subscripts in the conditions of the Proposition 5, we arrive at
the following fundamental statement.

\begin{theorem} 
Let $\g$ be equipped with the Poisson bracket {\rm PB($\A, \B, \C, \D$)}.
Suppose that the following relations hold:
\[
A_{i+1,j+1}=-B_{i+1,j}=C_{i,j+1}=-D_{i,j}\quad{\rm for}\quad i\neq j;
\]
\[
A_{j+1,j+1}-D_{j,j}+B_{j+1,j}-C_{j,j+1}=0\quad{\rm for\;\;all}\quad 1\le j\le n.
\]
Then each map
\begin{equation}
{\cal M}\circ \sigma^j :\g\mapsto g,\qquad {\cal M}\circ \sigma^j(\uu)=T_{j}
\end{equation}
is Poisson, if $g$ is equipped with the Poisson bracket
\[
{\rm PB(}A_{j+1,j+1}, B_{j+1,j}, C_{j,j+1}, D_{j,j}).
\] 
The Hamiltonian equations 
of motion generated by the Hamiltonian function {\rm(\ref{inv ham})} 
with an $Ad$--invariant function $\varphi$, are of the form
\begin{equation}
\dot{u}_j=u_j\,{\cal L}_{j-1}-{\cal L}_ju_j,\qquad
{\cal L}_j=R_j(d\varphi(T_j)),
\end{equation}
where 
\[
R_j=A_{j+1,j+1}+B_{j+1,j}=D_{j,j}+C_{j,j+1}.
\]
(In all the formulas the subscripts should be taken {\rm (mod $n$)}).
\end{theorem}

This Theorem may be considered as a final link in the chain of generalizations
\cite{STS1}, \cite{STS2}, \cite{LP1}, \cite{LP2}, \cite{L}, \cite{NCPQ}, 
\cite{STS4}, \cite{STSS}. However, for the non--ultralocal quadratic Poisson 
structures on direct sums discussed in these papers the operators $\A,\B,\C,\D$ 
always had only few nonvanishing ''operator entries'', namely $A_{j,j}$, 
$D_{j,j}$, $B_{j+1,j}$, and $C_{j,j+1}$. The ''most nonlocal'' example (though 
only with $n=2$) appeared in \cite{S} in connection with the relativistic Toda 
lattice.

\setcounter{equation}{0}
\section{Basic algebras and operators}
Two sorts of algebras will play the basic role in our presentation. They are
well suited to describe various lattice systems with the so called open--end
and periodic boundary conditions, respectively. 

1) For the {\it open--end case} we set $g=gl(N)$, the algebra of
$N\times N$ matrices 
\begin{equation}
u=\sum_{i,j=1}^N u_{ij}E_{ij}.
\end{equation}
The (nondegenerate, invariant) scalar product is choosen as
\begin{equation}
\langle u,v\rangle={\rm tr}(u\cdot v).
\end{equation}

2) For the {\it periodic case}  ${g}\subset gl(N)[\lambda,\lambda^{-1}]$ 
is a twisted subalgebra of the loop algebra $gl(N)[\lambda,\lambda^{-1}]$,
consisting of formal semiinfinite Laurent series over $gl(N)$ satisfying
\begin{equation}
u(\lambda)=
\sum_{p\in{\Bbb Z} \atop p<<\infty}
\sum_{i-j\equiv p \atop({\rm mod}\,N)}
\lambda^p u_{ij}^{(p)}E_{ij}.
\end{equation}
The scalar product is choosen as 
\begin{equation}
\langle u(\lambda),\,v(\lambda)\rangle={\rm tr}(u(\lambda)\cdot v(\lambda))_0,
\end{equation}
the subscript 0 denoting the free term of the formal Laurent series. 

In these cases of matrix algebras the Poisson bracket PB($A,B,C,D$) may be 
written in a fine tensor form.

1) In the {\it open--end case} functions $u_{ij}$ form the functional basis of 
the set of functions on $g$. It is easy to see that $\nabla u_{ij}=E_{ji}$. For 
a linear operator $R$ on $g$ define the corresponding $N^2\times N^2$ matrix 
$r$:
\begin{equation}\label{r open}
r=\sum_{i,j,k,l}\left\langle R(\nabla u_{ij}),\nabla u_{kl}\right\rangle
E_{ij}\otimes E_{kl}.
\end{equation}
Then it is easy to check that the pairwise Poisson brackets of the coordinate
functions may be cast into the formula
\begin{equation}\label{u,u open}
\{u\stackrel{\otimes}{,}u\}=(u\otimes u)\,a-d\,(u\otimes u)
+(I\otimes u)\,b\,(u\otimes I)-(u\otimes I)\,c\,(I\otimes u).
\end{equation}
Here the matrices $a,b,c,d$ correspond to the operators $A,B,C,D$ in the same
manner, as $r$ corresponds to $R$.

2) Analogously, in the {\it periodic case} the functional basis of the set of 
functions on $g$ is formed by the coordinates $u_{ij}^{(p)}$, for which
$\nabla u_{ij}^{(p)}=\lambda^{-p}E_{ji}$. The $n^2\times N^2$ matrix 
$r(\lambda,\mu)$ depending on two parameters $\lambda$, $\mu$, corresponding 
to a linear operator $R$, is now defined as
\begin{equation}\label{r per}
r(\lambda,\mu)=
\sum_{\renewcommand{\arraystretch}{0.5}
\begin{array}{c}\scriptstyle p,q\\\scriptstyle i-j\equiv p\;({\rm mod} N)\\
\scriptstyle k-l\equiv q\;({\rm mod}\,N)\end{array}}
\left\langle R\left(\nabla u_{ij}^{(p)}\right),\nabla u_{kl}^{(q)}\right\rangle
\lambda^p \mu^q E_{ij}\otimes E_{kl}.
\end{equation}
The pairwise Poisson brackets of the coordinate functions may be presented in 
the form analogous to (\ref{u,u open}):
\begin{eqnarray}
&& \{u(\lambda)\stackrel{\otimes}{,}u(\mu)\} \nonumber\\
&& =(u(\lambda)\otimes u(\mu))\,a(\lambda,\mu)
-d(\lambda,\mu)\,(u(\lambda)\otimes u(\mu))\nonumber\\
&& +(I\otimes u(\mu))\,b(\lambda,\mu)\,(u(\lambda)\otimes I)
-(u(\lambda)\otimes I)\,c(\lambda,\mu)\,(I\otimes u(\mu)).
\label{u,u per}
\end{eqnarray}

The Poisson brackets PB($\A,\B,\C,\D$) on the ''big'' algebra $\g$ may be
also presented in the tensor form. For example, in the {\it periodic case}
the tensor representation of the bracket (\ref{big PB}) reads:
\begin{eqnarray*}
&& \{u_j(\lambda)\stackrel{\otimes}{,}u_i(\mu)\} \\ 
&& =(u_j(\lambda)\otimes u_i(\mu))\,a_{ij}(\lambda,\mu)
-d_{ij}(\lambda,\mu)\,(u_j(\lambda)\otimes u_i(\mu))\\
&& +(I\otimes u_i(\mu))\,b_{ij}(\lambda,\mu)\,(u_j(\lambda)\otimes I)
-(u_j(\lambda)\otimes I)\,c_{ij}(\lambda,\mu)\,(I\otimes u_i(\mu)).
\label{u,u big}
\end{eqnarray*}
(In the {\it open--end case} one has simply to omit everywhere the spectral 
parameters $\lambda$, $\mu$).

The usefulness of the tensor notation lies in that it provides us with an 
efficient method of finding Poisson submanifolds for the bracket PB($A,B,C,D$).
Suppose that there is a Poisson space $S$ with a Poisson bracket 
$\{\;,\;\}_0$ and a local coordinates $s$, and a map $T(s):S\mapsto g$. Suppose 
also that the matrix $\{T(s)\stackrel{\otimes}{,}T(s)\}_0$ composed of pairwise 
Poisson brackets of entries of the matrix $T(s)$, may be presented in the form 
(\ref{u,u open}) or (\ref{u,u per}), respectively (naturally, in terms of $T(s)$ 
instead of $u$). Then the set $T(S)\subset g$ is a Poisson submanifold for the
bracket PB($A,B,C,D$). For a simple proof see \cite{S}.

Now we introduce several operators which will be widely used in the following
presentation.

1) In the {\it open--end case} $g$ as a linear space may be presented as a 
direct sum
\[
g=\bigoplus_{k=-N+1}^{N-1}g_k,
\]
where $g_k$ is the set of matrices with nonzero entries only on the
$k$th diagonal, i.e. in the positions $(i,j)$ with $i-j=k$. In particular,
$g_0$ is a set of diagonal matrices, which serves as a commutative
subalgebra of $g$. The other two obvious subalgebras are:
\[
{g}_+=\bigoplus_{k=1}^{M-1}{g}_k, \qquad
{g}_-=\bigoplus_{k=-M+1}^{-1}{g}_k,
\]
i.e. the sets of strictly lower and upper triangular matrices. 
 
2) In the {\it periodic case} $g$ as a linear space is a direct sum
\[
g=\bigoplus_{k\in{\Bbb Z}\atop k<<\infty} {g}_k,
\]
where ${g}_k$ is the set of monomial matrices from $g$ containing
only the power $\lambda^k$. Again ${g}_0$ is a commutative subalgebra
of $g$. The two other subalgebras are:
\[
{g}_+=\bigoplus_{0<k<<\infty}{g}_k, \qquad
{g}_-=\bigoplus_{k<0}{g}_k,
\]
i.e. the sets of Laurent series with positive and with negative powers of
$\lambda$, respectively.

In the both cases, ${g}$ as a linear space is a direct sum of three 
subspaces ${g}_+$, ${g}_-$, ${g}_0$, and we denote the 
corresponding projections by $P_+$, $P_-$, $P_0$, respectively. (Obviously,
as a set ${g}_0$ is identical in both the open--end and the periodic 
case, but, of course, $P_0$ has different meaning in these two cases).

Define a linear operator $R_0$ on ${g}$ by
\begin{equation}\label{R0}
R_0=P_+-P_-.
\end{equation}

Define also a skew--symmetric linear operator $\Pi$ on ${g}_0$ by
\begin{equation}\label{Pi}
\Pi(E_{jj})=\sum_{k=1}^N \pi_{jk}E_{kk},\quad \pi_{jk}=
\left\{\begin{array}{cl}1,& j>k\\-1,& j<k\\0,& j=k\end{array}\right.,
\end{equation}
and continue $\Pi$ on the whole $g$ according to $\Pi=\Pi\circ P_0$.

Then the following operators satisfy all conditions of the Proposition 1:
\begin{equation}\label{ABCD}
\begin{array}{ccc}
A=\frac{1}{2}(R_0+\Pi), & \quad & 
D=\frac{1}{2}(R_0-\Pi),\\ \\
B=\frac{1}{2}(P_0-\Pi), & \quad &
C=\frac{1}{2}(P_0+\Pi),
\end{array}
\end{equation}
so that the Poisson bracket PB($A, B, C, D$) is defined. Note that these 
operators have the property (\ref{sum}):
\begin{equation}\label{R}
A+B=C+D=R=\textstyle\frac{1}{2}(R_0+P_0).
\end{equation}

An important property of these operators was established in \cite{S3}.

\begin{proposition}
Fix an element ${\cal E}\in {g}_1$:
\[
{\cal E}=\sum_{k=1}^{N-1} E_{k+1,k} \quad {\rm or} \quad
{\cal E}=\lambda \sum_{k=1}^N E_{k+1,k}
\]
in the open--end and periodic case, respectively. Then for an arbitrary
natural number $m\ge 2$ the set
\begin{equation}\label{Pm}
{\cal P}_{m-1}={\cal E}\oplus\bigoplus_{j=0}^{m-1} {g}_{-j}
\end{equation}
is a Poisson submanifold for the bracket {\rm PB($A, B, C, D$)}. 
\end{proposition}

To conclude this section we give also the expressions for the matrices 
$r_0$, $p_0$, $\pi$ corresponding to the operators $R_0$, $P_0$, $\Pi$,
respectively. The expression for the $r_0$ depends on the case. In the
{\it open--end case}:
\begin{equation}\label{r0 open}
r_0=\sum_{i<j}E_{ij}\otimes E_{ji} - \sum_{i>j}E_{ij}\otimes E_{ji}.
\end{equation}
In the {\it periodic case}:
\begin{equation}\label{r0 per}
r_0(\lambda,\mu)=
\frac{\lambda^N+\mu^N}{\lambda^N-\mu^N}\sum_{i=1}^N E_{ii}\otimes E_{ii} 
+\sum_{p=1}^{N-1}\frac{2\lambda^p\mu^{N-p}}{\lambda^N-\mu^N}
\sum_{i-j\equiv p\atop({\rm mod}\,N)}E_{ij}\otimes E_{ji}.
\end{equation}
The expressions for $p_0$, $\pi$ are identical in the both cases:
\begin{equation}\label{p0}
p_0=\sum_{i=1}^N E_{ii}\otimes E_{ii},
\end{equation}
\begin{equation}\label{pi}
\pi=\sum_{i,j=1}^N \pi_{ij}E_{ii}\otimes E_{jj}.
\end{equation}

The matrices $a,b,c,d$ corresponding to the operators $A,B,C,D$ are given 
in the {\it periodic case} by
\begin{equation}\label{abcd}
\begin{array}{ccc}
a(\lambda,\mu)=\frac{1}{2}\Big(r_0(\lambda,\mu)+\pi\Big), & \; &
d(\lambda,\mu)=\frac{1}{2}\Big(r_0(\lambda,\mu)-\pi\Big),\\ \\
b=\frac{1}{2}(p_0-\pi), & \quad &
c=\frac{1}{2}(p_0+\pi),
\end{array}
\end{equation}
and in the {\it open--end case} one has to omit the spectral parameters
$\lambda$, $\mu$.

\setcounter{equation}{0}
\section{Bogoyavlensky lattice}
We will be studying the following integrable lattice system, known as
Bogoyavlensky lattice \cite{B} (although it was introduced earlier in 
\cite{N}, and its certain special case appeared also in \cite{I}):
\begin{equation}\label{BL}
\dot{z}_k=z_k\left(\sum_{j=1}^{m-1}z_{k+j}-\sum_{j=1}^{m-1}z_{k-j}\right).
\end{equation}

It may be considered on an infinite lattice (all the subscripts 
belong to ${\Bbb Z}$), and admits also finite--dimensional reductions of two 
types:open--end and periodic. We shall denote the number of lattice cites
(particles) in the finite reductions by $M-m+1$ in the open--end case, and by
$M$ in the periodic case, for the reasons which will become clear soon.
The boundary conditions in the open--end case are: 
\[
\quad z_k=0\;\;{\rm for}\;\; k\le 0\;\;{\rm and\;\;for}\;\;k\ge M-m+2;
\]
in the periodic case all the subscripts belong to ${\Bbb Z}/M{\Bbb Z}$.

A Hamiltonian interpretation of this system was found in \cite{B}: the system
(\ref{BL}) is Hamiltonian with the Hamiltonian function
\begin{equation}\label{H}
H(z)=\sum z_k
\end{equation}
with respect to the Poisson bracket given by:
\begin{equation}\label{PB}
\{z_j,z_k\}=\left\{\begin{array}{rl}z_jz_k,&\quad j-k=1,\ldots, m-1\\
                                    -z_jz_k,&\quad j-k=-m+1,\ldots, -1\\
                                     0,&\quad {\rm else}
                    \end{array}\right.
\end{equation}
(in the periodic case with $M>2m-2$ the conditions of the type
$j-k=1,\ldots, m-1$ have to be understood (mod $M$)). 

Bogoyavlensky has found also the Lax representations for this system:
\begin{equation}\label{Lax}
\dot{T}=\left[T,P\right],
\end{equation}
where 
\begin{equation}\label{T}
T(z,\lambda)=\lambda^{-m+1}\sum z_kE_{k,k+m-1}+\lambda\sum E_{k+1,k},
\end{equation}
\begin{equation}\label{P}
P(z,\lambda)=\sum(z_k+z_{k-1}+\ldots+z_{k-m+1})E_{kk}+\lambda^{m}\sum 
E_{k+m,k},
\end{equation}

Here for the infinite lattices all the subscripts belong to ${\Bbb Z}$,
for the periodic case all the subscripts belong to ${\Bbb Z}/M{\Bbb Z}$,
and for the open--end case all the subscripts belong to ${1,\ldots,M}$
(so that for the both types of finite reductions the matrices involved in
the Lax representations are $M\times M$).
Moreover, in the infinite--dimensional and open--end cases the dependence
on the spectral parameter $\lambda$ becomes inessential and may be suppressed
by setting $\lambda=1$. Below we consider only finite lattices. It can be said
that in these cases the Lax matrix $T$ belongs to the algebra $\goth g$,
which is defined exactly as $g$ in the previous section with the only
difference: $N$ should be everywhere replaced by $M$.

In \cite{S3} we gave an $r$--matrix interpretation of the Bogoyavlensky
lattices. The main results of \cite{S3} concerning the lattice (\ref{BL})
may be summarized as follows. Let the operators $A,B,C,D$ on $\goth g$ be
defined as in the previous section, with $N$ replaced by $M$. Define also a
Posson bracket PB($A,B,C,D$) on $\goth g$. Then, according to the Proposition
7, the set ${\cal P}_{m-1}\subset {\goth g}$ defined as in (\ref{Pm}), is a 
Poisson submanifold. Obviously, the set ${\cal T}_{m-1}$ consisting of the
Lax matrices (\ref{T}), is a subset of ${\cal P}_{m-1}$.

\begin{proposition} The Dirac reduction of the bracket {\rm PB($A,B,C,D$)} 
to the subset ${\cal T}_{m-1}\subset{\cal P}_{m-1}$ coincides with the 
bracket {\rm(\ref{PB})}. The Dirac correction to the Hamiltonian vector field 
$\{\varphi,u\}$, when reducing it to the set ${\cal T}_{m-1}$, vanishes, if 
$\varphi(u)=\psi(u^{m})$ with some $Ad$--invariant function $\psi$.
\end{proposition}

This Proposition explains the Lax equation (\ref{Lax}), because the Hamiltonian
function (\ref{H}) can obviously be seen as
\[
H=\frac{1}{m}\,{\rm tr}(T^{m}),\quad{\rm so\;\; that}\;\; dH=d\,'H=T^{m},
\]
and it is not difficult to check that for the matrix (\ref{P}) there holds:
\[
P=(P_++P_0)(T^{m})=\left(R+\textstyle\frac{1}{2}I\right)(T^{m}).
\]

\setcounter{equation}{0}
\section{Illustrative example: Volterra lattice}

We intend now to establish another Lax representation for the Bogoyavlensky 
lattices, alternative to (\ref{Lax}). To illustrate our approach by the
simple particular case, we start with the lattice (\ref{BL}) corresponding 
to $m=2$, and known also as the Volterra (or Lotka--Volterra, or Langmuir)
lattice:
\begin{equation}\label{VL}
\dot{z}_k=z_k(z_{k+1}-z_{k-1}),
\end{equation}
The invariant Poisson bracket (\ref{PB}) for this system is:
\begin{equation}\label{VL PB}
\{z_{k+1},z_k\}=z_{k+1}z_k,
\end{equation}
and the corresponding Hamiltonian function
\begin{equation}\label{VL H}
H(z)=\sum_k z_k.
\end{equation}

We want to separate all the particles into  two ''sorts'':
\begin{equation}\label{uv}
u_k=z_{2k-1},\quad v_k=z_{2k}.
\end{equation}
In terms of these two sorts of particles the equations of motion (\ref{VL})
may be put down as
\begin{equation}\label{VL in uv}
\dot{u}_k=u_k(v_k-v_{k-1}),\qquad \dot{v}_k=v_k(u_{k+1}-u_k).
\end{equation}
The underlying Poisson bracket (\ref{VL PB}) is now
\begin{equation}\label {VL PB in uv}
\{v_k,u_k\}=v_ku_k,\qquad \{u_{k+1},v_k\}=u_{k+1}v_k,
\end{equation}
and the Hamiltonian function (\ref{VL H}) is equal to
\begin{equation}\label{VL H in uv}
H(u,v)=\sum_k (u_k + v_k).
\end{equation}

\subsection{Periodic case}
The notation (\ref{uv}) is consistent with the $M$--periodic boundary 
conditions, only if the total number of particles is {\it even}. Hence we 
consider in this subsection only the case
\begin{equation}\label{M=2N}
M=2N.
\end{equation}

The usual $M\times M=2N\times 2N$ Lax matrix for the Volterra lattice is:
\begin{equation}\label{VL T}
T(z,\lambda)=\lambda\sum_{k=1}^{2N} E_{k+1,k}
+\lambda^{-1}\sum_{k=1}^{2N}z_kE_{k,k+1}.
\end{equation}
Its natural ambient space is the algebra $\goth g$ defined in the previous 
Section.

Our aim in this Section is to elaborate an alternative $N\times N$ Lax 
representation for this system, naturally living in the algebra $g$ whose 
definition was given in the Section 5.

Define two Lax matrices from $g$:
\begin{equation}\label{VL L1}
L_1(u,v,\lambda)=\lambda\sum_{k=1}^N E_{k+1,k}+\sum_{k=1}^N(u_k+v_{k-1})E_{kk}
+\lambda^{-1}\sum_{k=1}^Nu_kv_kE_{k,k+1}.
\end{equation}
\begin{equation}\label{VL L2}
L_2(u,v,\lambda)=\lambda\sum_{k=1}^N E_{k+1,k}+\sum_{k=1}^N(u_k+v_k)E_{kk}
+\lambda^{-1}\sum_{k=1}^Nu_{k+1}v_kE_{k,k+1},
\end{equation}
\begin{proposition} 

1) The set of matrices $L_1(u,v,\lambda)$ (or $L_2(u,v,\lambda)$) forms
a Poisson submanifold in $g$, if the latter is equipped with the Poisson
bracket {\rm PB($A, B, C, D$)}.

2) The Volterra lattice {\rm(\ref{VL in uv})} admits two (equivalent) Lax 
representations
\begin{equation}\label{VL alt Lax}
\dot{L}_1=[L_1,R(L_1)],\qquad \dot{L}_2=[L_2,R(L_2)].
\end{equation}

3) The function
\begin{equation}\label{det:L=L}
\det\Big(L_1(u,v,\lambda)-\mu I_N\Big)=\det\Big(L_2(u,v,\lambda)-\mu I_N\Big)
\end{equation}
serves as a generating function of integrals of motion of the Volterra lattice.
Moreover,
\begin{equation}\label{det:T=LL}
\det\Big(T^2(z,\lambda)-\mu I_M)=
\det\Big(L_1(u,v,\lambda^2)-\mu I_N\Big)\det\Big(L_2(u,v,\lambda^2)-\mu I_N\Big).
\end{equation}
\end{proposition}
{\bf Proof.} We prove first of all the third statement, which will also give
a motivation for considering the matrices (\ref{VL L1}), (\ref{VL L2}).

The generating function of integrals of motion for the periodic Volterra 
lattice, following from the $M\times M$ Lax representation, can be chosen as
\begin{equation}\label{VL gf}
\det(T^2(z,\lambda)-\mu I).
\end{equation}
Here the matrix $T^2(z,\lambda)$ has the following structure:
\begin{equation}\label{VL T2}
T^2(z,\lambda)=\lambda^2\sum_{k=1}^{2N} E_{k+2,k}
+\sum_{k=1}^{2N}(z_k+z_{k-1})E_{kk}
+\lambda^{-2}\sum_{k=1}^{2N}z_kz_{k+1}E_{k,k+2}.
\end{equation}
We use the Laplace formula to represent the determinant (\ref{VL gf}) as
\begin{eqnarray}
\det(T^2-\mu I) & = & \sum_{i_1<\ldots<i_N}
\det(T^2-\mu I)\left(\begin{array}{cccc} 1 & 3 & \ldots & 2N-1\\
i_1 & i_2 & \ldots & i_N\end{array}\right) \nonumber \\                   
 & \times & \det(T^2-\mu I)\left(\begin{array}{cccc} 2 & 4 & \ldots & 2N\\
j_1 & j_2 & \ldots & j_N\end{array}\right)
\label{VL Laplace}
\end{eqnarray}
Here the sum is taken over all possible ordered $N$--tuples 
$(i_1,\ldots,i_N)$ of the natural numbers from $(1,\ldots,2N)$, and
the numbers $j_1<\ldots<j_N$ form the complement of 
$(i_1,\ldots,i_N)$ to $(1,\ldots,2N)$. We use the notation
\[
{\cal A}\left(\begin{array}{cccc} 
\alpha_1 & \alpha_2 & \ldots & \alpha_N\\
\beta_1 & \beta_2 & \ldots & \beta_N   
              \end{array}\right)
\]
for the submatrix of $\cal A$ formed by the elements standing on the 
intersection of the rows $(\alpha_1,\ldots,\alpha_N)$ with the columns 
$(\beta_1,\ldots,\beta_N)$.

Now the crucial observation based on the explicit formula (\ref{VL T2})
and on the hypothesis (\ref{M=2N}) is the following: in fact in the huge
sum (\ref{VL Laplace}) all but one terms vanish identically, so that we are
left with
\begin{eqnarray}
\det(T^2-\mu I) & = & 
\det(T^2-\mu I)\left(\begin{array}{cccc} 1 & 3 & \ldots & 2N-1\\
1 & 3 & \ldots & 2N-1\end{array}\right) \nonumber \\                   
 & \times & \det(T^2-\mu I)\left(\begin{array}{cccc} 2 & 4 & \ldots & 2N\\
2 & 4 & \ldots & 2N\end{array}\right).  
\label{VL T^2}
\end{eqnarray}
But it is easy to calculate that
\begin{equation}\label{VL T^2 to L1}
(T^2(z,\lambda)-\mu I_{2N})\left(\begin{array}{cccc} 1 & 3 & \ldots & 2N-1\\
1 & 3 & \ldots & 2N-1\end{array}\right)=L_1(u,v,\lambda^2)-\mu I_N,
\end{equation}
\begin{equation}\label{VL T^2 to L2}
(T^2(z,\lambda)-\mu I_{2N})\left(\begin{array}{cccc} 2 & 4 & \ldots & 2N\\
2 & 4 & \ldots & 2N\end{array}\right)=L_2(u,v,\lambda^2)-\mu I_N.
\end{equation}
This proves the formula (\ref{det:T=LL}). To prove the formula (\ref{det:L=L}),
notice that the matrices $L_1$, $L_2$ are in fact connected by means of a
similarity transformation, due to the following {\it important observation}:
\begin{equation}
L_1(u,v,\lambda)=U(u,\lambda)V(v,\lambda),\qquad 
L_2(u,v,\lambda)=V(v,\lambda)U(u,\lambda),
\end{equation}
where
\begin{equation}\label{VL wU}
U(u,\lambda)=\lambda\sum_{k=1}^N E_{k+1,k}+\sum_{k=1}^N u_kE_{kk},
\end{equation}
\begin{equation}\label{VL V}
V(v,\lambda)=I+\lambda^{-1}\sum_{k=1}^N v_kE_{k,k+1}.
\end{equation}
The third statement of the Proposition is herewith completely proved.

Turning to the first statement, we notice that the both matrices $L_1$, $L_2$
have the same structure as the Lax matrix of the Toda lattice (cf. \cite{S}):
\begin{equation}\label{TL L}
L(a,b,\lambda)=\lambda\sum_{k=1}^N E_{k+1,k}+\sum_{k=1}^Nb_kE_{kk}
+\lambda^{-1}\sum_{k=1}^Na_kE_{k,k+1}.
\end{equation}
The expressions for the ''Toda coordinates'' $(a,b)$ are given for the
matrix $L_1$ by
\begin{equation}\label{VL to TL 1}
b_k=u_k+v_{k-1},\qquad a_k=u_kv_k,
\end{equation}
and for the matrix $L_2$ by
\begin{equation}\label{VL to TL 2}
b_k=u_k+v_k,\qquad a_k=u_{k+1}v_k.
\end{equation}

Now a very remarkable circumstance is that in the both parametrizations 
(\ref{VL to TL 1}), (\ref{VL to TL 2}) the equations of motion (\ref{VL in uv})
imply the equations of motion of the Toda lattice:
\begin{equation}\label{TL}
\dot{b}_k=a_k-a_{k-1},\qquad \dot{a}_k=a_k(b_{k+1}-b_k),
\end{equation}
while the Poisson brackets (\ref{VL PB in uv}) imply the {\it quadratic}
Poisson brackets for the Toda lattice:
\begin{equation}\label{TL PB}
\begin{array}{rcl}
\{b_{k+1},b_k\}=a_k,  & \quad & \{a_{k+1},a_k\}=a_{k+1}a_k, \\
\{b_k,a_k\}=-b_ka_k, & \quad & \{b_{k+1},a_k\}=b_{k+1}a_k.
\end{array}
\end{equation}
The corresponding Hamiltonian function (\ref{VL H in uv}) in the both 
parametrizations is equal to
\begin{equation}\label{TL H}
H=\sum_{k=1}^N b_k.
\end{equation}

An $r$--matrix interpretation of the quadratic Poisson bracket (\ref{TL PB})
for the Toda lattice was found in \cite{S}. It was demonstrated there that the 
matrices $L(a,b,\lambda)$ form a Poisson submanifold in the algebra $g$ equipped
with PB($A,B,C,D$). The proof in \cite{S} consisted of a verification of the 
tensor form of this result:
\begin{eqnarray}
\{L(\lambda)\stackrel{\otimes}{,}L(\mu)\}=
\Big(L(\lambda)\otimes L(\mu)\Big)\,a(\lambda,\mu)
-d(\lambda,\mu)\,\Big(L(\lambda)\otimes L(\mu)\Big) &&\nonumber\\
+\Big(I\otimes L(\mu)\Big)\, b\,\Big(L(\lambda)\otimes I\Big)
-\Big(L(\lambda)\otimes I\Big)\, c\, \Big(I\otimes L(\mu)\Big).&&
\label{L,L}
\end{eqnarray}

In the formula (\ref{L,L}) $L$ stands for the Lax matrix (\ref{TL}) of the 
Toda lattice, but it might as well stand for either of the Lax matrices 
(\ref{VL L1}), (\ref{VL L2}) of the Volterra lattice. 
This already proves the first statement of the Proposition. Taking into account 
that the Hamiltonian function (\ref{VL H in uv}) is equal to
\[
H=\varphi(L_1)=\varphi(L_2),
\]
where
\[
\varphi(L)={\rm tr}(L)_0\quad{\rm so\;\;that}\quad d\varphi(L)=L,
\]
we see that the second statement of the Proposition is also proved.\qed

However, we want to re-derive the basic result (\ref{L,L}) here from the 
viewpoint of the monodromy maps, which will simplify its proof and provide 
an additional information.

\begin{proposition} 
The pairs of matrices $(U,V)$ form a Poisson submanifold in the algebra
$\g=g\oplus g$ equipped with the Poisson bracket PB($\A,\B,\C,\D$), where
\[
\A=\left(\begin{array}{cc} A & -B\\ C & A\end{array}\right), \qquad
\B=\left(\begin{array}{cc} B & B\\ B & -C\end{array}\right),
\]
\[
\C=\left(\begin{array}{cc} C & C\\ C & -B\end{array}\right), \qquad
\D=\left(\begin{array}{cc} D & -C\\ B & D\end{array}\right).
\]
\end{proposition}
{\bf Proof.} The proof of this statement consists of the direct verification
of the following identities (each of them is simpler then (\ref{L,L}), because
they deal only with bidiagonal matrices, while (\ref{L,L}) -- with tridiagonal
ones):
\begin{eqnarray*}
&&\{U(\lambda)\stackrel{\otimes}{,}U(\mu)\} \\
&&=\Big(U(\lambda)\otimes U(\mu)\Big)\,a(\lambda,\mu)
-d(\lambda,\mu)\,\Big(U(\lambda)\otimes U(\mu)\Big) \\
&& +\Big(I\otimes U(\mu)\Big)\,b\,\Big(U(\lambda)\otimes I\Big)
-\Big(U(\lambda)\otimes I\Big)\,c\,\Big(I\otimes U(\mu)\Big),\\
&&\\
&&\{V(\lambda)\stackrel{\otimes}{,}V(\mu)\} \\
&&=\Big(V(\lambda)\otimes V(\mu)\Big)\,a(\lambda,\mu)
-d(\lambda,\mu)\,\Big(V(\lambda)\otimes V(\mu)\Big) \\
&& -\Big(I\otimes V(\mu)\Big)\, c\,\Big(V(\lambda)\otimes I\Big)
+\Big(V(\lambda)\otimes I\Big)\, b\, \Big(I\otimes V(\mu)\Big),\\
&&\\
&&\{U(\lambda)\stackrel{\otimes}{,}V(\mu)\} \\
&&=\Big(U(\lambda)\otimes V(\mu)\Big)\,c
-b\,\Big(U(\lambda)\otimes V(\mu)\Big) \\
&& +\Big(I\otimes V(\mu)\Big)\, b\,\Big(U(\lambda)\otimes I\Big)
-\Big(U(\lambda)\otimes I\Big)\, c\, \Big(I\otimes V(\mu)\Big),\\
&&\\
&&\{V(\lambda)\stackrel{\otimes}{,}U(\mu)\} \\
&&=-\Big(V(\lambda)\otimes U(\mu)\Big)\,b
+c\,\Big(V(\lambda)\otimes U(\mu)\Big) \\
&& +\Big(I\otimes U(\mu)\Big)\, b\,\Big(V(\lambda)\otimes I\Big)
-\Big(V(\lambda)\otimes I\Big)\, c\, \Big(I\otimes U(\mu)\Big).
\end{eqnarray*}
(Of course, the last two identities are equivalent, so that it is enough 
to verify one of them).\qed
\vspace{5mm}

Now the conditions of the Theorem 1 with $n=2$ read:
\[
A_{12}=-B_{11}=C_{22}=-D_{21},\quad A_{21}=-B_{22}=C_{11}=-D_{12},
\]
\[
A_{22}-D_{11}+B_{21}-C_{12}=0,
\]
\[
A_{11}-D_{22}+B_{12}-C_{21}=0,
\]
and are, obviously, fulfilled. This Theorem assures that the monodromy
map $\g\mapsto g$ defined as $(U,V)\mapsto V\cdot U=L_2$ is Poisson, if the 
target space $g$ is equipped with the Poisson bracket 
\[
{\rm PB(}A_{11}, B_{12},C_{21},D_{22})={\rm PB(}A,B,C,D),
\] 
and that the map $(U,V)\mapsto U\cdot V=L_1$ is also Poisson, if the target 
space $g$ is equipped with the Poisson bracket
\[
{\rm PB(}A_{22}, B_{21},C_{12},D_{11})={\rm PB(}A,B,C,D), 
\]
which coincides with the previous one. 
It follows that the manifold consisting of the matrices $L_2(u,v,\lambda)$ 
(or of the matrices $L_1(u,v,\lambda)$) is 
Poisson in this bracket, as an image of a Poisson manifold under a Poisson map.

The Theorem 1 implies also that the Hamiltonian equations of motion
on the pairs $(U,V)$ with the Hamiltonian function 
\[
\varphi(UV)=\varphi(VU)\qquad(\varphi\;\;Ad-{\rm invariant})
\]
may be presented in the form of the ''Lax triads''
\[
\dot{U}=U\cdot R(d\varphi(VU))-R(d\varphi(UV))\cdot U,
\]
\[
\dot{V}=V\cdot R(d\varphi(UV))-R(d\varphi(VU))\cdot V,
\]
where 
\[
R=A+B=C+D=A_{11}+B_{12}=A_{22}+B_{21}=D_{11}+C_{12}=D_{22}+C_{21}.
\]
These Lax triads imply also the Lax equations for the matrices $L_1=VU$, 
$L_2=UV$:
\[
\dot{L}_1=\Big[L_1,R(d\varphi(L_1))\Big], \quad
\dot{L}_2=\Big[L_2,R(d\varphi(L_2))\Big].
\]

We would like to notice that these results are parallel to those obtained 
in \cite{S} for the relativistic Toda lattice.

\subsection{Open--end case}
We outline briefly the features of the open--end case different from that of
the periodic one.

First of all, the condition (\ref{M=2N}) is no more necessary for the
applicability of our construction. However, it remains more elegant in this
casebeing completely parallel to periodic case. We consider this variant first,
i.e. the Volterra lattice with open ends and {\it odd} number of particles:
\[
u_k,\;\;k=1,\ldots,N,\quad{\rm and}\quad v_k,\;\;k=1,\ldots,N-1.
\]
The both Lax matrices $L_1$, $L_2$ belong to one and the same algebra $g$,
the $N\times N$ open--end version of the algebra $\goth g$:
\begin{equation}\label{open L1}
L_1(u,v)=\sum_{k=1}^{N-1} E_{k+1,k}+\sum_{k=1}^N(u_k+v_{k-1})E_{kk}
+\sum_{k=1}^{N-1}u_kv_kE_{k,k+1}.
\end{equation}
\begin{equation}\label{open L2}
L_2(u,v)=\sum_{k=1}^{N-1} E_{k+1,k}+\sum_{k=1}^N(u_k+v_k)E_{kk}
+\sum_{k=1}^{N-1}u_{k+1}v_kE_{k,k+1},
\end{equation}
(In the first of these formulas $v_0=0$, in the second one $v_N=0$).
All the statements of the Proposition 10 remain valid, if one drops out
the dependence on the spectral parameter $\lambda$. Also the factorization
\begin{equation}\label{open fact}
L_1(u,v)=U(u)V(v),\quad L_2(u,v)=V(v)U(u)
\end{equation}
holds, with the matrices
\[
U(u)=\sum_{k=1}^{N-1}E_{k+1,k}+\sum_{k=1}^N u_kE_{kk}, \quad
V(v)=I+\sum_{k=1}^{N-1}v_kE_{k,k+1}.
\]
As in the periodic case, the pairs $(U,V)$ form a Poisson submanifold in
$g\oplus g$ carrying the correspondent Poisson bracket.

Consider now the variant with
\[
M=2N-1,
\]
when the open--end Volterra lattice consists of an {\it even} number of 
particles:
\[
u_k,\;\;k=1,\ldots,N-1,\quad{\rm and}\quad v_k,\;\;k=1,\ldots,N-1,
\]
and the usual Lax matrix is $2N-1\times 2N-1$. In this case the matrices 
$L_1$, $L_2$ have {\it different dimensions}, namely $L_1$ is still 
$N\times N$ and is given by (\ref{open L1}) (this time with vanishing $u_N$),
while $L_2$ is $N-1\times N-1$:
\begin{equation}\label{open L2'}
L_2(u,v)=\sum_{k=1}^{N-2} E_{k+1,k}+\sum_{k=1}^{N-1}(u_k+v_k)E_{kk}
+\sum_{k=1}^{N-2}u_{k+1}v_kE_{k,k+1}.
\end{equation}
Correspondingly, the two alternative Lax representations of the Volterra 
lattice, analogous to (\ref{VL alt Lax}), are still valid, but live in 
{\it two different algebras}; the Lax matrices still form Poisson submanifolds, 
but also in {\it two different algebras}; their characteristic polynomials
do not coincide any more, but satisfy instead the identity
\[
\det\Big(L_1(u,v)-\mu I_{N}\Big)=(-\mu)\det\Big(L_2(u,v)-\mu I_{N-1}\Big).
\]
An interpretation in terms of the Poisson bracket on $g\oplus g$ fails
in this formulation. 

All these inconveniences, however, can be repaired if we include the Volterra
lattice with $2N-2$ particles into the one with $2N-1$ particles, i.e. 
add a dummy particle $u_{N}$ with a trivial evolution: $u_{N}=0$. Then
the both matrices $L_1$, $L_2$ become $N\times N$ and coincide with 
(\ref{open L1}), (\ref{open L2}) ($L_2$ having vanishing dummy entries 
in the positions $(N,N)$ and $(N-1,N)$), and the whole construction (including 
the factorization (\ref{open fact}) and the Poisson property of the manifold 
$(U,V)$) remains valid. Indeed, the condition $u_{N}=0$ is compatible not only 
with the equations of motion (\ref{VL in uv}), but also with the Poisson 
brackets (\ref{VL PB in uv}).

\setcounter{equation}{0}
\section{Alternative Lax representation \newline
for the general Bogoyavlensky lattice}

\subsection{Periodic case}
For the general Bogoyavlensky lattice (\ref{BL}) we separate all the particles 
into $m$ sorts, according to:
\begin{equation}
v_k^{(j)}=z_{m(k-1)+j},\quad j=1,2,\ldots,m.
\end{equation}
This is consistent with the $M$--periodic boundary conditions, only if
\begin{equation}\label{M=mN}
M=mN.
\end{equation}
In these new notations the equations of motion (\ref{BL}) take the form:
\begin{equation}\label{BL in v}
\dot{v}_k^{(j)}=v_k^{(j)}\left(\sum_{i=j+1}^m\Big(v_k^{(i)}-v_{k-1}^{(i)}\Big)
+\sum_{i=1}^{j-1}\Big(v_{k+1}^{(i)}-v_k^{(i)}\Big)\right).
\end{equation}
The quadratic Poisson bracket (\ref{PB}) invariant under this flow, may be 
presented in the following nice form:
\begin{equation}\label{PB in v}
\{v_k^{(i)},v_k^{(j)}\}=v_k^{(i)}v_k^{(j)},\quad 
\{v_{k+1}^{(j)},v_k^{(i)}\}=v_{k+1}^{(j)}v_k^{(i)}\quad
{\rm for}\quad 1\leq j<i\leq m.
\end{equation}
(As usual, all other brackets vanish).
The corresponding Hamiltonian function (\ref{H}) is equal to
\begin{equation}\label{H in v}
H=\sum_{k=1}^N\sum_{j=1}^m v_k^{(j)}.
\end{equation}

Introduce the following $N\times N$ matrices, depending on the variables
corresponding to the particles of only one sort, and on the spectral 
parameter $\lambda$:
\begin{equation}\label{V j}
V_j(\lambda)=I+\lambda^{-1}\sum_{k=1}^N v_k^{(j)}E_{k,k+1},\quad
j=1,2,\ldots,m.
\end{equation}
We suppress in this notation the argument $v^{(j)}$ of the matrix $V_j$,
because it can be restored unambigously from the subscript.
We shall need also the matrix
\begin{equation}\label{U 1}
U_1(\lambda)=\lambda\sum_{k=1}^N E_{k+1,k}+\sum_{k=1}^N v_k^{(1)}E_{kk}.
\end{equation}

Now we can define the Lax matrix as a ''monodromy matrix'':
\begin{equation}\label{Lm}
L_m(v^{(1)},\ldots,v^{(m)},\lambda)=
V_m(\lambda)\cdot\ldots\cdot V_2(\lambda)\cdot U_1(\lambda)
\end{equation}
and its shifted versions as
\begin{equation}\label{Lj}
L_j(v^{(1)},\ldots,v^{(m)},\lambda)=V_j(\lambda)\cdot\ldots\cdot
V_2(\lambda)\cdot U_1(\lambda)\cdot V_m(\lambda)\cdot\ldots\cdot
V_{j+1}(\lambda).
\end{equation}

\begin{theorem}

1) The set of matrices $L_j(v^{(1)},\ldots,v^{(m)},\lambda)$ for each
$1\le j\le m$ forms a Poisson submanifold in the algebra $g$ equipped 
with the Poisson bracket {\rm PB($A,B,C,D$)}.

2) The Bogoyavlensky lattice {\rm(\ref{BL in v})} admits a set of $m$
(equivalent) Lax representations:
\begin{equation}
\dot{L}_j=[L_j,R(L_j)].
\end{equation}

3) The generating function of integrals of motion in the $N\times N$ 
representation is connected with an analogous object in the $M\times M$
representation by
\begin{equation}
\det\Big(T^m(z,\lambda)-\mu I_M)
=\prod_{j=1}^m\det\Big(L_j(v^{(1)},\ldots,v^{(m)},\lambda^m)-\mu I_N\Big),
\end{equation}
all factors in this product being mutually equal.
\end{theorem}
{\bf Proof.} As by the proof of the Proposition 8, we start with the third 
statement. Apply the Laplace formula to expand the determinant
\[
\det\Big(T^m(z,\lambda)-\mu I_M\Big),
\]
according to the decomposition of the rows into $m$ complementary sets, the 
$j$th of them ($1\le j\le m$) having the numbers $(j,m+j,\ldots,m(N-1)+j)$.
The matrix $T^m$ has a very special structure, namely 
\begin{equation}\label{T^m}
T^m(z,\lambda)\in{\cal E}^m+\bigoplus_{i=0}^{m-1}{\goth g}_{-mi}
\end{equation}
(cf. (\ref{VL T2}) for $m=2$). Taking this structure into account, we see that 
only one from the huge number of terms in the Laplace formula does not vanish:
\[
\det\Big(T^m-\mu I_M\Big)=\prod_{j=1}^m
\det\Big(T^m-\mu I_M\Big)\left(\begin{array}{cccc}
j & m+j & \ldots & m(N-1)+j \\j & m+j & \ldots & m(N-1)+j \end{array}\right).
\]
Now the submatrices of $T^m$ can be directly calculated, which gives:
\begin{eqnarray}
&&\Big(T^m(z,\lambda)-\mu I_M\Big)\left(\begin{array}{cccc}
j & m+j & \ldots & m(N-1)+j \\j & m+j & \ldots & m(N-1)+j \end{array}\right)=
\nonumber\\   \nonumber\\ 
&&=L_j(v^{(1)},\ldots,v^{(m)},\lambda^m)-\mu I_N
\label{T^m to Lj}
\end{eqnarray}
(cf. for $m=2$ the calculation of the submatrices (\ref{VL T^2 to L1}), 
(\ref{VL T^2 to L2}) resulting in the expressions (\ref{VL L1}), (\ref{VL L2})).

Since all matrices $L_j(v^{(1)},\ldots,v^{(m)},\lambda^m)$ are connected by
means of a similarity transformation, their determinants are mutually equal, 
which proves the third statement of the Proposition.

The first statement follows from the Proposition 7. Indeed, it is easy to see 
from (\ref{T^m to Lj}), (\ref{T^m}) that the set
of matrices $L_j(v^{(1)},\ldots,v^{(m)},\lambda)$ for each $j$ is exactly the 
\[
{\cal P}_{m-1}={\cal E}\oplus\bigoplus_{i=0}^{m-1} {g}_{-i}.
\]
The second statement follows now from the general $r$--matrix theory, because
the function (\ref{H in v}) may be presented as $H=\varphi(L_j)$, where
$\varphi(L)={\rm tr}(L)_0$, so that $d\varphi(L)=L$. \qed

We want, however, just as in the previous Section, to give another proof of 
the first statement of the Theorem 2, based upon
the interpretation of the matrices $L_j(v^{(1)},\ldots,v^{(m)},\lambda)$ as
monodromy matrices. 

\begin{theorem} The ordered $m$--tuples $(U_1,V_2,\ldots,V_m)$ form a
Poisson submanifold in the algebra $\g=\bigoplus_{j=1}^m g$, if the latter is
equipped with the Poisson bracket {\rm PB($\A,\B,\C,\D$)}, where the operators 
$\A, \B, \C, \D$ are defined according to the formulas:
\begin{eqnarray*}
A_{ij} & = & \left\{\begin{array}{rl}
A, &\;\;{\rm if}\;\; i=j\\
-B,&\;\;{\rm if}\;\;i=1,j>1\;\;{\rm or}\;\;i>j>1\\
C, &\;\;{\rm if}\;\;i>1,j=1\;\;{\rm or}\;\;j>i>1
\end{array}\right.\\ \\
B_{ij} & = & \left\{\begin{array}{rl} 
B, &\;\;{\rm if}\;\; i=1\;\;{\rm or}\;\;i>j\\
-C,&\;\;{\rm if}\;\;j\ge i>1
\end{array}\right.\\ \\
C_{ij} & = & \left\{\begin{array}{rl} 
-B, &\;\;{\rm if}\;\; i\ge j>1\\
-C,&\;\;{\rm if}\;\;j>i\;\;{\rm or}\;\;j=1
\end{array}\right.\\ \\
D_{ij} & = & \left\{\begin{array}{rl} 
D, &\;\;{\rm if}\;\; i=j\\
B, &\;\;{\rm if}\;\;i>j\\
-C,&\;\;{\rm if}\;\;j>i
\end{array}\right.
\end{eqnarray*}                        
\end{theorem}
{\bf Proof.} To prove this statement, one may straightforwardly verify the
following identities:
\begin{eqnarray*}
&&\{U_1(\lambda)\stackrel{\otimes}{,}U_1(\mu)\} \\
&&=\Big(U_1(\lambda)\otimes U_1(\mu)\Big)\,a(\lambda,\mu)
-d(\lambda,\mu)\,\Big(U_1(\lambda)\otimes U_1(\mu)\Big) \\
&& +\Big(I\otimes U_1(\mu)\Big)\,b\,\Big(U_1(\lambda)\otimes I\Big)
-\Big(U_1(\lambda)\otimes I\Big)\,c\,\Big(I\otimes U_1(\mu)\Big),\\
&&\\
&&\{V_j(\lambda)\stackrel{\otimes}{,}V_j(\mu)\} \\
&&=\Big(V_j(\lambda)\otimes V_j(\mu)\Big)\,a(\lambda,\mu)
-d(\lambda,\mu)\,\Big(V_j(\lambda)\otimes V_j(\mu)\Big) \\
&& -\Big(I\otimes V_j(\mu)\Big)\, c\,\Big(V_j(\lambda)\otimes I\Big)
+\Big(V_j(\lambda)\otimes I\Big)\,b\,\Big(I\otimes V_j(\mu)\Big),\\
&&\\
&&\{U_1(\lambda)\stackrel{\otimes}{,}V_j(\mu)\} \\
&&=\Big(U_1(\lambda)\otimes V_j(\mu)\Big)\,c
-b\,\Big(U_1(\lambda)\otimes V_j(\mu)\Big) \\
&& +\Big(I\otimes V_j(\mu)\Big)\, b\,\Big(U_1(\lambda)\otimes I\Big)
-\Big(U_1(\lambda)\otimes I\Big)\,c\,\Big(I\otimes V_j(\mu)\Big)
\quad(j>1),\\
&&\\
&&\{V_j(\lambda)\stackrel{\otimes}{,}U_1(\mu)\} \\
&&=-\Big(V_j(\lambda)\otimes U_1(\mu)\Big)\,b
+c\,\Big(V_j(\lambda)\otimes U_1(\mu)\Big) \\
&& +\Big(I\otimes U_1(\mu)\Big)\, b\,\Big(V_j(\lambda)\otimes I\Big)
-\Big(V_j(\lambda)\otimes I\Big)\,c\,\Big(I\otimes U_1(\mu)\Big)
\quad(j>1),\\
&&\\
&&\{V_i(\lambda)\stackrel{\otimes}{,}V_j(\mu)\} \\
&&=-\Big(V_i(\lambda)\otimes V_j(\mu)\Big)\,b
-b\,\Big(V_i(\lambda)\otimes V_j(\mu)\Big) \\
&& +\Big(I\otimes V_j(\mu)\Big)\, b\,\Big(V_i(\lambda)\otimes I\Big)
+\Big(V_i(\lambda)\otimes I\Big)\,b\,\Big(I\otimes V_j(\mu)\Big)
\quad(i>j),\\
&&\\
&&\{V_i(\lambda)\stackrel{\otimes}{,}V_j(\mu)\} \\
&&=\Big(V_i(\lambda)\otimes V_j(\mu)\Big)\,c
+c\,\Big(V_i(\lambda)\otimes V_j(\mu)\Big) \\
&& -\Big(I\otimes V_j(\mu)\Big)\, c\,\Big(V_i(\lambda)\otimes I\Big)
-\Big(V_i(\lambda)\otimes I\Big)\,c\,\Big(I\otimes V_j(\mu)\Big)
\quad(i<j).
\end{eqnarray*}
(The third of these identities is equivalent to the fourth one, and the fifth
is equivalent to the sixth. Therefore in fact one needs to verify only four 
identities). Propositions 3, 4 allow to check easily that the operators
$\A,\B,\C,\D$ satisfy the conditions of the Proposition 1 and thus indeed
define a Poisson bracket PB($\A,\B,\C,\D$).  \qed 
\vspace{5mm}

Now a careful inspection convinces that all the conditions of the Theorem 1 
with $n=m$ zre fulfilled. This Theorem states that the maps
\[
(U_1,V_2,\ldots,V_m)\mapsto L_j
\]
are Poisson, if the target spaces carry the Poisson brackets
\[
{\rm PB}(A_{j+1,,j+1},B_{j+1,j},C_{j,j+1},D_{j,j})={\rm PB}(A,B,C,D).
\]
Together with the Theorem 3 this implies the Poisson property of the
manifold formed by the matrices $L_j$ with respect to the latter bracket.

Further, it follows from the Theorem 1 that the equations of motion of
an arbitrary flow of the Bogoyavlensky lattice hierarchy with the Hamiltonian
function $\varphi(L)$ ($L+L_j$, $\varphi$ $Ad$--invariant) admits a 
Lax representation:
\begin{eqnarray*}
\dot{U}_1 & = & U_1\cdot R(d\varphi(L_m)) - R(d\varphi(L_1))\cdot U_1,\\
\dot{V}_j & = & V_j\cdot R(d\varphi(L_{j-1})) - R(d\varphi(L_j))\cdot V_j, 
\quad 2\le j\le m.
\end{eqnarray*}
As a consequence, the following Lax equations hold:
\[
\dot{L}_j=[L_j,R(d\varphi(L_j))],\quad 1\le j\le m.
\]
The Bogoyavlensky lattice proper corresponds here to $\varphi(L)=L$, 
$d\varphi(L)=L$.

\subsection{Open--end case}
As for the Volterra lattice ($m=2$), the Theorem 2 holds almost literally in 
the open--end case, if
\[
M=mN,
\]
i.e. for the lattice consisting of $m(N-1)+1$ particles
\[
v_k^{(1)},\;\; 1\le k\le N;\quad {\rm and}\quad v_k^{(j)},\;\;1\le k\le N-1 
\quad {\rm for}\quad j=2,\ldots,m.
\]
All that has to be changed, is to omit the spectral parameter $\lambda$ in
all formulas and to define the matrices $U_1$, $V_j$ as
\[
U_1=\sum_{k=1}^{N-1} E_{k+1,k}+\sum_{k=1}^N v_k^{(1)}E_{kk},
\]
\[
V_j=I+\sum_{k=1}^{N-1}v_k^{(j)}E_{k,k+1}.
\]

For the lattices with the number of particles different from $m(N-1)+1$ we 
can add one or several dummy particles with the trivial dynamics, so that
all the results remain valid.

There is, however, one more problem concerning the open--end case that has to
be mentioned, namely the problem of lacking integrals of motion. Indeed, the
generating function
\[
\det(L_j-\mu I_N)
\]
gives now only only $N$ integrals of motion, which for $m>2$ is much less
than necessary for complete integrability. The lacking integrals in a closely
related problem (full Toda lattice) were constructed in \cite{DLNT}, \cite{EFS}.
Analogous construction can be performed also for the both Lax representations
($M\times M$ and $N\times N$) for the Bogoyavlensky lattice, leading
presumably to two sets of additional integrals. However, our argument based
on the Laplace formula, does not imply the coincidence or even some relations
between these two sets. It would be very important to study this problem in
detail.

\section{Concluding remarks}
In the present paper a well known integrable lattice system was further
studied, which gave an opportunity to touch two general problems of the
theory of integrable systems, more precisely, of the Hamiltonian aspects 
of this theory. The first of these problems was completely solved here,
namely it was found the most general conditions for monodromy map on a direct
sum of associative algebras to be Poisson with respect to some general 
quadratic brackets. We were able to get rid of all sorts of locality
conditions, and to provide an example where such nonlocal structures
naturally arise.

Another general problem, a problem of {\it duality} between different
Lax representations for one and the same system, remains almost completely
open. Our present results merely add one new example of this situation,
and we hope that its careful analysis will bring us more close to the solution
of the general problem.

The research of the author is financially supported by the DFG (Deutsche
Forschungsgemeinschaft).

\end{document}